\documentclass{JFM-FLM_Au}

\lefttitle{F. Lyu, L. Zhao, W. Huang and C. Xu}
\righttitle{Journal of Fluid Mechanics}

\title{Log--linear law of the mean streamwise velocity in turbulent boundary layers with moderate adverse pressure gradients}

\author{Fuzhou Lyu\aff{1},
 Lihao Zhao\aff{1},  
 Weixi Huang\aff{1},  
 \and Chunxiao Xu\aff{1}}

\affiliation{\aff{1}AML, Department of Engineering Mechanics, Tsinghua University, Beijing 100084, People's Republic of China}

\corresau{Chunxiao Xu, \email{xucx@tsinghua.edu.cn}}

\begin{document}
\maketitle

\begin{abstract}
An essential feature of canonical zero-pressure-gradient (ZPG) turbulent boundary layers (TBLs) is that the mean streamwise velocity exhibits a logarithmic dependence on the wall-normal distance, known as the log law. In this study, we demonstrate that this conventional log law is not suitable for turbulent boundary layers subjected to pressure gradients (PGs). Instead, a log--linear law is theoretically derived for TBLs under moderate adverse pressure gradients (APGs), based on the total shear-stress balance and a rescaled eddy-viscosity model. The validity of the proposed log--linear law is assessed using available datasets of incompressible APG TBLs with the Clauser pressure-gradient parameter $\beta$ ranging from 0.73 to 9.0. Compared with the conventional log law, the present log--linear formulation shows significantly improved agreement with the measured mean velocity profiles. In the limiting case of $\beta \to 0$, the proposed law naturally recovers the classical log law.
\end{abstract}



\section{Introduction}
\label{sec:Introduction}
TBLs are classical flow configurations that have been extensively studied for decades due to their practical applications and theoretical importance. Most of the previous studies of TBL focus on the canonical ZPG cases. On the other hand, compared with ZPG TBLs, TBLs with pressure gradients are more common in practical applications such as aircraft wings and turbine blades. Due to the advances of experimental and numerical techniques, studies of PG TBLs have become increasingly active in the past decade. However, some fundamental issues remain unresolved, especially regarding the applicability of the conventional log law to PG TBLs.

For canonical ZPG TBLs, the log law states that the mean streamwise velocity $U$ in the overlap region between the inner and outer layers follows a logarithmic dependence on the wall-normal distance $y$, i.e.,
\begin{equation}
  U^+_{log} = \frac{1}{\kappa} \ln y^+ + B,
  \label{eq:log_law}
\end{equation}
where the superscript $+$ denotes normalization by wall units, defined using the friction velocity $u_\tau$ and the viscous length scale $\delta_\nu$, with
$u_\tau = \sqrt{\tau_w / \rho}$ and $\delta_\nu = \nu / u_\tau$.
Here, $\tau_w$ is the wall shear stress, and $\rho$ and $\nu$ are the fluid density and kinematic viscosity, respectively.
The K\'arm\'an constant is typically taken as $\kappa \approx 0.41$, while $B \approx 5.0$ denotes the intercept of the log law. This simple equation laid the foundation for understanding wall-bounded turbulent flows \citep{Pope2000,white2006viscous}, and has been widely employed in Reynolds-averaged Navier-Stokes (RANS) models \citep{wilcox1998turbulence} and wall-modeled large eddy simulation (WMLES) \citep{larsson2016large}. The region where the log law holds is usually referred to as the log region. Here, we use the term ``intermediate region'' to refer to the conventional log region, as we will show later that the log law is not applicable to APG TBLs. 

When pressure gradients are included, the flow properties become more complex. In particular, it has been recognized that the mean velocity profiles in the intermediate region of APG TBLs shift below the log law of ZPG TBLs \citep{nagano1998structure,devenport2022equilibrium} and the slope of the log law changes with the pressure gradient \citep{skaare1994turbulent}. Traditionally, these deviations are attributed to the variation of the K\'{a}rm\'{a}n constant $\kappa$. For instance, \citet{nickels2004inner}, \citet{nagib2008variations}, \citet{chauhan2010empirical}, and \citet{monty2011parametric} all reported lower values of $\kappa$ for APG TBLs compared to the standard value of 0.41 for ZPG TBLs. In addition, studies by \citet{huang1995law}, \citet{monty2011parametric} and \citet{lee2017large} show that the wall-normal range where the log law holds is also reduced for APG TBLs, even when the K\'{a}rm\'{a}n constant has been modified. These issues pose challenges to the analysis and modeling of APG TBLs. For example, \citet{crook2002skin} summarized that a $\pm0.5$ change in $1/\kappa$ results in a $12\%$ difference in $u_\tau$, which corresponds to an uncertainty of approximately $25\%$ in the friction coefficient. Furthermore, the reduction of the effective range of the log law makes WMLES sensitive to the choice of the matching layer, as this layer is usually required to be placed in the log region \citep{baggett1997some}. Therefore, it is necessary to reevaluate the applicability of the log law in APG TBLs and develop a more general velocity law that can better capture the behavior of APG TBLs.

The objective of the present study is to develop such a velocity law that is more robust than the conventional log law and explicitly accounts for pressure gradient effects. As will be shown later, both the apparent variation of the K\'{a}rm\'{a}n constant $\kappa$ and the reduction in the effective range of the log law can be attributed to the inapplicability of the classical log law to APG turbulent boundary layers. To address this limitation, a new log--linear law is proposed. Compared with the conventional log law, the proposed formulation exhibits a substantially wider range of validity, while the K\'{a}rm\'{a}n constant $\kappa$ retains its classical value of 0.41. Within this framework, the classical log law emerges as a special case of the log--linear law in the absence of a pressure gradient.

\section{log--linear law for APG TBLs}
For canonical ZPG TBLs, the log law can be derived from the eddy viscosity hypothesis and the total-shear-stress equation, leading to
\begin{equation}
  \frac{\nu_t}{\nu} \frac{\partial U^+}{\partial y^+} = \tau_t^+,
  \label{eq:total_shear_stress}
\end{equation}
where $\nu_t$ is the eddy viscosity. In the intermediate layer, it is commonly assumed that
\begin{equation}
  \nu_t/\nu = \kappa y^+.
  \label{eq:nut}
\end{equation}
In Eq.~(\ref{eq:total_shear_stress}), the left-hand side represents the Reynolds shear stress under the eddy viscosity assumption, while the viscous stress is neglected. The right-hand side is the total shear stress $\tau_t^+$. In the near-wall region of ZPG TBLs, $\tau_t^+ \approx 1$ is a good approximation. Therefore, substituting $\tau_t^+ \approx 1$ and $\nu_t/\nu = \kappa y^+$ into Eq.~(\ref{eq:total_shear_stress}) and then integrating with respect to $y^+$ yields the log law. Equation~(\ref{eq:total_shear_stress}) is derived from the momentum equation and is thus applicable to APG TBLs. However, the other two assumptions, namely $\tau_t^+ \approx 1$ and $\nu_t/\nu = \kappa y^+$, need to be revisited in the context of pressure gradients.

To quantitatively assess the strength of the pressure gradient, the Clauser pressure gradient parameter $\beta = (\delta_d / \tau_w)(\mathrm{d}P_e/\mathrm{d}x)$ is considered, where $\delta_d$ is the displacement thickness, $x$ is the streamwise coordinate, and $P_e$ is the pressure at the edge of the boundary layer \citep{clauser1954turbulent}. This parameter has been extensively used in studies of APG TBLs \citep{bobke2017history,devenport2022equilibrium,wei2023outer}. Systematic comparisons by \citet{monty2011parametric} also show that $\beta$ consistently collapses APG flow statistics and thus serves as an effective descriptor of PG effects.

We first reconsider the assumption $\tau_t^+ \approx 1$. The Reynolds-averaged momentum equation for turbulent boundary layers with pressure gradients reads
\begin{equation}
  U \frac{\partial U}{\partial x} + V \frac{\partial U}{\partial y} = -\frac{1}{\rho} \frac{\mathrm{d} P_e}{\mathrm{d} x} - \frac{\partial \overline{u'v'}}{\partial y} + \nu \frac{\partial^2 U}{\partial y^2},
  \label{eq:momentum_equation}
\end{equation}
where $V$ is the mean wall-normal velocity and $\overline{u'v'}$ is the Reynolds shear stress. In the case of strong pressure gradients where the flow tends to separate, the Reynolds normal stresses neglected in Eq.~(\ref{eq:momentum_equation}) become significant \citep{simpson1977features}, and thus the present analysis is limited to moderate APGs. 
Integrating Eq.~(\ref{eq:momentum_equation}) in the near-wall region and neglecting the convective terms yields the following total shear stress relation:
\begin{equation}
  \frac{\tau_t}{\rho} = \nu \frac{\partial U}{\partial y} - \overline{u'v'} \approx \left( 1 + \beta \frac{y}{\delta_d} \right) \frac{\tau_w}{\rho}.
  \label{eq:tau_t}
\end{equation}
Thus, $\tau_t^+$ is no longer constant in the near-wall region, but instead varies linearly with the wall-normal distance due to PG effects.

In the presence of pressure gradients, the assumption $\nu_t/\nu = \kappa y^+$ may no longer hold either. The wall coordinate $y^+$ is defined as the wall distance normalized by the viscous length scale $\delta_\nu$. Since $\delta_\nu$ depends entirely on the wall shear stress $\tau_w$, the normalization based on $\delta_\nu$ is appropriate for ZPG TBLs, where the total shear stress $\tau_t$ is approximately constant in the near-wall region. However, for TBLs with pressure gradients, the distribution of $\tau_t$ is strongly influenced by PG effects, as observed by \citet{perry1994wall}. Therefore, there is no compelling reason to retain the assumption $\nu_t/\nu = \kappa y/\delta_\nu$.
An alternative approach is to consider the K\'{a}rm\'{a}n momentum integral equation. For PG TBLs, this equation can be written as
\begin{equation}
  \frac{\mathrm{d}}{\mathrm{d}x}(U_e^2 \theta)
  = u_\tau^2(1+\beta),
  \label{eq:karman}
\end{equation}
where $\theta$ is the momentum thickness and the subscript $e$ denotes quantities at the edge of the boundary layer. According to Eq.~(\ref{eq:karman}), a new velocity scale and the corresponding length scale can be defined as
\begin{equation}
  \begin{aligned}
  u_\beta &= u_\tau \sqrt{1+\beta}, \\
  \delta_\beta &= \frac{\nu}{u_\beta}.
  \end{aligned}
  \label{eq:new_scale}
\end{equation}
With these definitions, Eq.~(\ref{eq:karman}) can be rewritten as $\mathrm{d}(U_e^2 \theta)/\mathrm{d}x = u_\beta^2$, which has the same form as that for ZPG TBLs. This set of scales has been employed to model the mean profiles of APG TBLs by \citet{shu2025mean}. Replacing $\delta_\nu$ in Eq.~(\ref{eq:nut}) with $\delta_\beta$ yields
\begin{equation}
  \frac{\nu_t}{\nu} = \kappa y^+ \sqrt{1+\beta}.
  \label{eq:nu_t}
\end{equation}
Substituting Eqs.~(\ref{eq:nu_t}) and (\ref{eq:tau_t}) into Eq.~(\ref{eq:total_shear_stress}) gives
\begin{equation}
  \kappa y^+ \sqrt{1+\beta} \frac{\partial U^+}{\partial y^+} = 1 + \beta \frac{y}{\delta_d}.
  \label{eq:total_shear_stress_refined}
\end{equation}
Integrating Eq.~(\ref{eq:total_shear_stress_refined}) with respect to $y^+$ yields the refined law for the mean streamwise velocity:
\begin{equation}
U^+ = \frac{1}{\sqrt{1+\beta}} \frac{1}{\kappa} \ln y^+ + \frac{\beta}{\sqrt{1+\beta}} \frac{y^+}{\kappa \delta_d^+} + C,
\label{eq:log_linear_law}
\end{equation}
where $C$ is an integration constant. In contrast to Eq.~(\ref{eq:log_law}), Eq.~(\ref{eq:log_linear_law}) contains a linear contribution in addition to the logarithmic term, and is therefore referred to as the log--linear law. Furthermore, when $\beta = 0$, Eq.~(\ref{eq:log_linear_law}) reduces to the log law in Eq.~(\ref{eq:log_law}). Thus, the present log--linear law can be viewed as a generalization of the conventional log law for APG TBLs.

\section{Validation of the proposed law}
To assess the validity of the proposed log--linear law, APG TBL cases from four research groups are considered, as summarized in table~\ref{tab:apg_tbl_cases}. All datasets are obtained from direct numerical simulation (DNS) or high-fidelity large-eddy simulation (LES). The cases are named according to the following convention: the prefix `B' denotes cases with a nearly constant Clauser parameter, where the subsequent number indicates the approximate value of $\beta$. For cases where $\beta$ varies significantly along the streamwise direction, the prefix `M' is used, followed by a number representing the exponent of the power-law free-stream velocity distribution $U_\infty \propto (x-x_0)^m$ applied at the upper boundary. This identifier is followed by the first letter of the first author's surname and the year of publication. For instance, B10B2017 corresponds to the case with $\beta \approx 1.0$ from \citet{bobke2017history}. For cases B10B2017--B14P2022, data are available at multiple streamwise stations, and the ranges of momentum thickness Reynolds number $Re_\theta$, friction Reynolds number $Re_\tau$, and Clauser parameter $\beta$ are provided. For B07L2017--B90L2017, only one flow section is available for each case, and the corresponding flow parameters are also listed.
\begin{table}
  \begin{center}
\def~{\hphantom{0}}
  \begin{tabular}{lccccl}
      Case       & $Re_\theta$   & $Re_\tau$  & $\beta$   & Data type & References \\[3pt]
      B10B2017   & 1561--2876    & 400--700   & 0.9--1.1  & LES       & \cite{bobke2017history}\\
      B20B2017   & 1792--3357    & 400--700   & 2.0--2.2  & LES       & \cite{bobke2017history}\\
      M13B2017   & 1991--3561    & 450--750   & 0.9--1.3  & LES       & \cite{bobke2017history}\\
      M16B2017   & 2104--3583    & 450--750   & 1.8--2.7  & LES       & \cite{bobke2017history}\\
      M18B2017   & 2564--3493    & 450--750   & 3.1--4.5  & LES       & \cite{bobke2017history}\\
      B14Y2018   & 2100--5700    & 531--803   & 1.3--1.5  & DNS       & \cite{yoon2018contribution}\\
      B14P2022   & 4808--8359    & 1000--1800 & 1.2--1.6  & LES       & \cite{pozuelo2022adverse}\\
      B07L2017   & 1605          & 364        & 0.73      & DNS       & \cite{lee2017large}\\
      B22L2017   & 2180          & 346        & 2.2       & DNS       & \cite{lee2017large}\\
      B90L2017   & 2840          & 318        & 9.0       & DNS       & \cite{lee2017large}\\
  \end{tabular}
\caption{Summary of the incompressible APG TBL datasets used in the present study. Cases B10B2017, B20B2017, M13B2017, M16B2017, and M18B2017 correspond to cases b1, b2, m13, m16, and m18, respectively, in table~2 of \cite{bobke2017history}; case B14Y2018 corresponds to the APG TBL case in table~1 of \cite{yoon2018contribution}; case B14P2022 corresponds to case b1.4 in table~1 of \cite{pozuelo2022adverse}; cases B07L2017, B22L2017, and B90L2017 correspond to cases in table~2 of \cite{lee2017large}.}
  \label{tab:apg_tbl_cases}
  \end{center}
\end{table}

\begin{figure}
\centering
\includegraphics[width=1\textwidth]{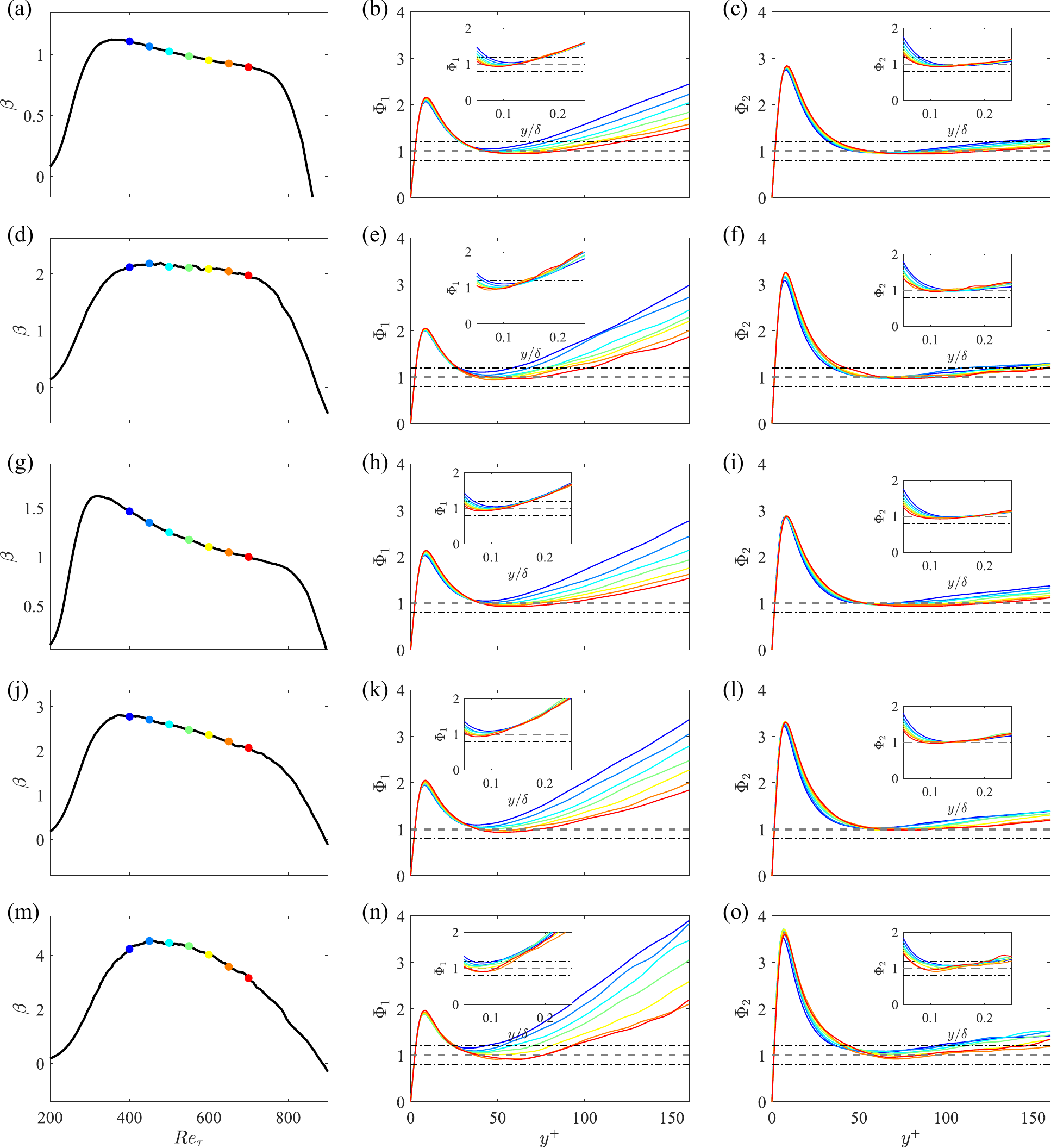}
\caption{ Flow configurations and diagnostic functions for case B10B2017 (a--c), B20B2017 (d--f), M13B2017 (g--i), M16B2017 (j--l), and M18B2017 (m--o).
The flow configurations, characterized by the Clauser parameter $\beta$ and friction Reynolds number $Re_\tau$, are presented by the markers in the left column. 
The diagnostic functions for the log law ($\Phi_{1}$) and the log--linear law ($\Phi_{2}$) are presented in the middle and right columns, respectively. 
Zoomed views of the diagnostic functions are provided in the insets in the middle and right columns. The horizontal grey dashed line indicates the plateau of $\Phi \approx 1$, with two black dash-dotted lines representing a tolerance of $\pm 20\%$ of the plateau.}
\label{fig:1}
\end{figure}

\begin{figure}
\centering
\includegraphics[width=1\textwidth]{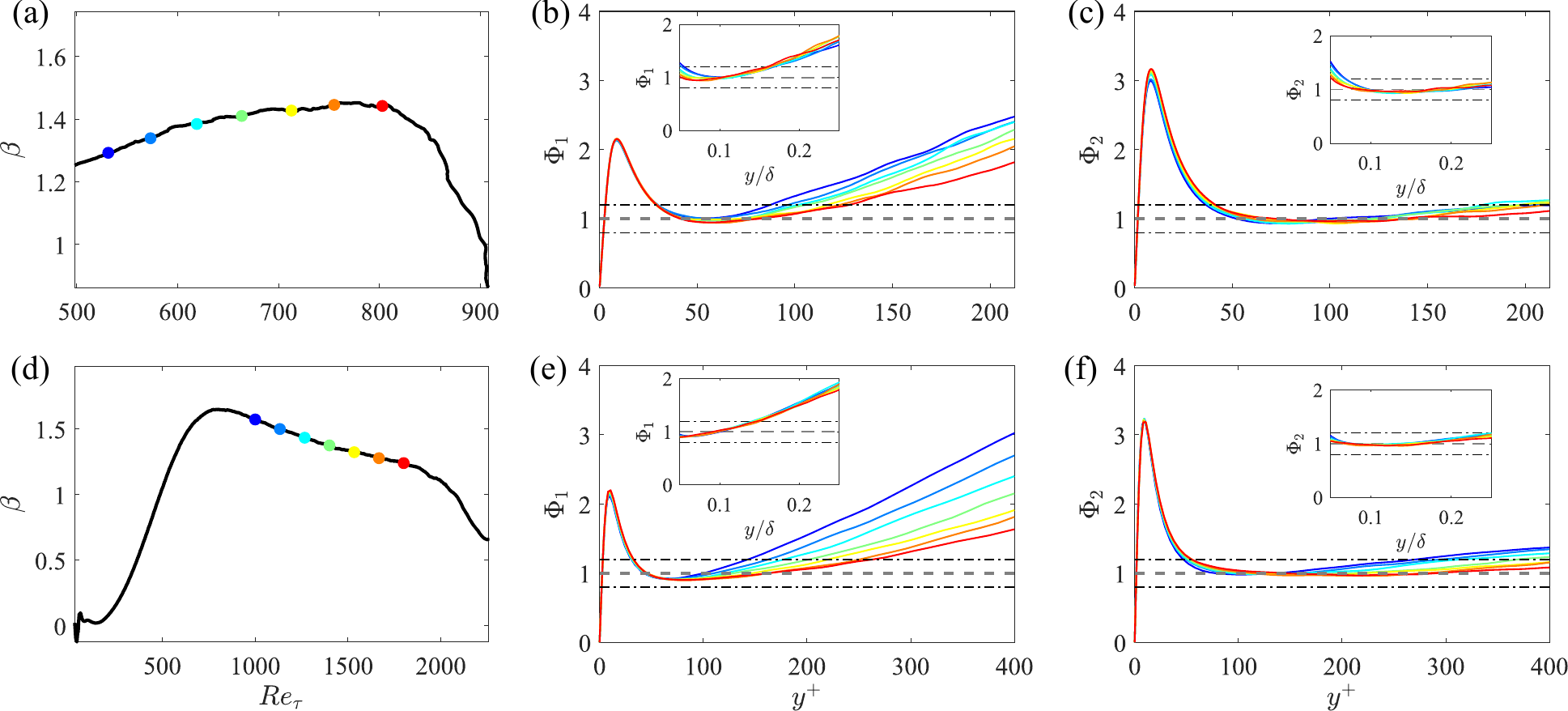}
\caption{Flow configurations and diagnostic functions for cases B14Y2018 (a--c) and B14P2022 (d--f). The flow configurations, characterized by the Clauser parameter $\beta$ and friction Reynolds number $Re_\tau$, are presented by the markers in the left column. The diagnostic functions for the log law ($\Phi_{1}$) and the log--linear law ($\Phi_{2}$) are presented in the middle and right columns, respectively. Zoomed views of the diagnostic functions are provided in the insets. The horizontal grey dashed line indicates the plateau of $\Phi \approx 1$, with two black dash-dotted lines representing a tolerance of $\pm 20\%$ regarding the plateau.}
\label{fig:2}
\end{figure}

\begin{figure}
\centering
\includegraphics[width=1\textwidth]{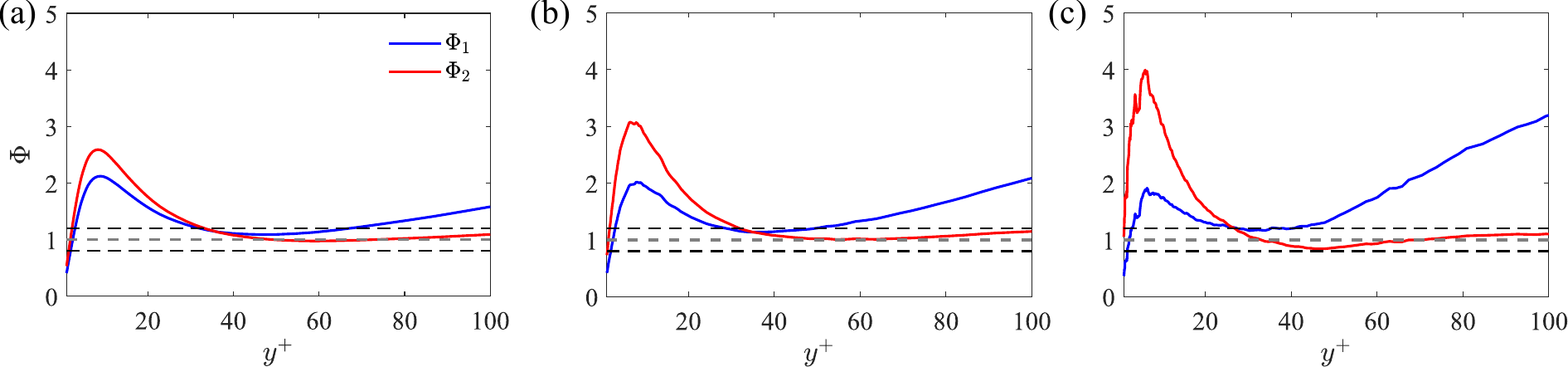}
\caption{Diagnostic functions for cases B07L2017 (a), B22L2017 (b), and B90L2017 (c). Blue lines and red lines represent the diagnostic functions for the log law and the log--linear law, respectively. The horizontal grey dashed line indicates the plateau of $\Phi \approx 1$, with two black dash-dotted lines representing a tolerance of $\pm 20\%$ regarding the plateau.}
\label{fig:3}
\end{figure}

\begin{figure}
\centering
\includegraphics[width=0.6\textwidth]{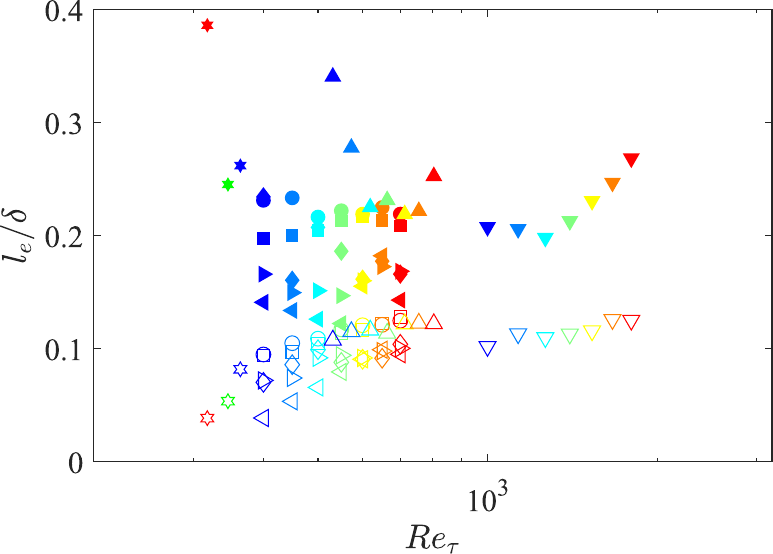}
\caption{Effective lengths for different cases. Circle, diamond, square, right-triangle, left-triangle, up-triangle, and down-triangle markers correspond to cases B10B2017, B20B2017, M13B2017, M16B2017, M18B2017, B14Y2018, and B14P2022, respectively. The marker colors correspond to the flow configurations in figures~\ref{fig:1} and \ref{fig:2}. The blue, green, and red hexagrams correspond to cases B07L2017, B22L2017, and B90L2017, respectively. The open and filled markers represent the effective lengths of the log law and the log--linear law, respectively.}
\label{fig:4}
\end{figure}

To validate the proposed log--linear law, we define the following diagnostic functions:
\begin{equation}
\begin{array}{rcl}
  \Phi_{1} & = & \displaystyle \kappa y^+ \frac{\partial U^+}{\partial y^+}, \\[8pt]
  \Phi_{2} & = & \displaystyle \frac{\kappa y^+ \sqrt{1+\beta}}{1+\beta y/\delta_d}
    \frac{\partial U^+}{\partial y^+},
\end{array}
\label{eq:diag_logs}
\end{equation}
where $\Phi_{1}$ and $\Phi_{2}$ are the diagnostic functions for the conventional log law and the present log--linear law, respectively. This form of diagnostic function has been widely used in previous studies \citep{monty2011parametric,pirozzoli2014revisiting}. A plateau of $\Phi \approx 1$ indicates that the corresponding law is satisfied. 

The diagnostic functions for cases B10B2017--M18B2017 are shown in figure~\ref{fig:1}, along with the distribution of $\beta$ with respect to $Re_\tau$. These cases share similar $Re_\tau$ but different $\beta$ values. Only the flow configurations at the middle streamwise stations are considered to avoid inflow and outflow effects. These configurations correspond to the markers in the left column of figure~\ref{fig:1}. As shown in the middle column of figure~\ref{fig:1}, the diagnostic function $\Phi_1$ rapidly deviates from 1 in the intermediate regions of the flow. Although the plateau region where $\Phi_1 \approx 1$ appears to extend with the streamwise development, the deviation trend remains unchanged when the wall distance is scaled by the boundary layer thickness, as evidenced by the zoomed views in the middle column. At the same $Re_\tau$, with an increase in $\beta$, the deviation of $\Phi_1$ from 1 becomes more pronounced. For example, at $Re_\tau = 517$, the diagnostic function $\Phi_1$ for case B10B2017 with $\beta = 1.0$ exceeds 1.2 at $y^+=91$ in the intermediate region, while $\Phi_1$ for case M18B2017 with $\beta = 4.4$ exceeds 1.2 at $y^+=79$. By contrast, the diagnostic function $\Phi_2$ shows much better agreement with the plateau of 1 in the intermediate regions and is insensitive to pressure gradient effects. 

The same behavior can be observed for cases B14Y2018 and B14P2022, as shown in figure~\ref{fig:2}. These two cases exhibit similar $\beta$ but different $Re_\tau$. For the diagnostic function $\Phi_1$, it can be found that at the same $\beta$, with the increase of $Re_\tau$, the deviation of both $\Phi_1$ and $\Phi_2$ from 1 becomes less pronounced. For example, at $\beta = 1.44$, $\Phi_1$ for case B14Y2018 ($Re_\tau=755$) exceeds 1.2 at $y^+=123$, while $\Phi_1$ for case B14P2022 ($Re_\tau=1267$) exceeds 1.2 at $y^+=178$. On the other hand, the diagnostic function $\Phi_2$ exceeds 1.2 at $y^+=209$ and $y^+=317$ for the same configurations. The Reynolds number effects are to be expected, as the upper boundary of the traditional log region for ZPG TBLs is also proportional to $Re_\tau$ \citep{smits2011high}. Generally speaking, the plateau of $\Phi_2 \approx 1$ is constantly wider than that of $\Phi_1 \approx 1$ for both cases considered here.

For cases B07L2017--B90L2017, where the Reynolds number is relatively low but $\beta$ spans a wide range from 0.73 to 9.0, the superiority of the proposed log--linear law is also evident, as shown in figure~\ref{fig:3}. When the pressure gradient is mild, the difference between $\Phi_1$ and $\Phi_2$ is relatively less distinct, as shown in figure~\ref{fig:3}(a). However, with an increase in $\beta$, the deviation of $\Phi_1$ from 1 becomes pronounced, while $\Phi_2$ maintains a reasonable plateau region of $\Phi_2\approx 1$, as shown in figures~\ref{fig:3}(b) and (c).

To quantitatively evaluate the superiority of the proposed log--linear law, we define the effective region of the log law and the log--linear law as the range where the corresponding diagnostic function remains within a tolerance of $\pm 20\%$ of the plateau $\Phi \approx 1$. The span of this region is referred to as the effective length $l_e$. As shown in figure~\ref{fig:4}, the effective lengths of the conventional log law are around $0.1\delta$ for most cases, whereas the effective lengths of the log--linear law are consistently larger, with an average value of $0.2\delta$. Specifically, for the case with the strongest pressure gradients, $\beta=9.0$ (hexagram marker in figure~\ref{fig:4}), the effective length of the log law is $0.04\delta$, while that of the log--linear law is $0.39\delta$, indicating a nearly 10-fold increase. Therefore, the proposed log--linear law significantly extends the effective region compared to the conventional log law.

\section{Conclusions}
In this study, we propose a more general velocity law, Eq.~(\ref{eq:log_linear_law}), for the intermediate region of APG TBLs, referred to as the log--linear law. In the limit of zero pressure gradient, the log--linear law reduces to the classical log law. Therefore, the log--linear law can be viewed as a generalization of the conventional log law for APG TBLs. This newly proposed law incorporates a linear term in addition to the logarithmic term. The pressure gradient effects are explicitly accounted for by considering the total shear stress equation and the rescaling of the eddy-viscosity model. It is shown that the log--linear law is superior to the conventional log law, significantly extending the effective range. These results suggest that previous observations of the apparent variation of the K\'{a}rm\'{a}n constant $\kappa$ and the reduction of the effective range of the log law in APG TBLs can be attributed to the inapplicability of the classical log law. Within the framework of the log--linear law, the K\'{a}rm\'{a}n constant $\kappa \approx 0.41$ remains unchanged regardless of the pressure gradient. Since the present log--linear law does not involve any additional empirical coefficients, it can be easily implemented in the modeling and prediction of APG TBLs.

\bibliographystyle{jfm}
\bibliography{jfm}

\end{document}